\documentclass[a4paper,fleqn,usenatbib]{mnras}

\usepackage{newtxtext,newtxmath}

\usepackage[T1]{fontenc}
\usepackage{ae,aecompl}

\usepackage{graphicx}	
\usepackage{amsmath}	
\usepackage{amssymb}	
\usepackage{caption}
\usepackage{booktabs}
\usepackage{subcaption}
\usepackage{multirow}
\usepackage{threeparttablex}  
\usepackage{dcolumn}
\usepackage{float}
\usepackage{pdflscape}
\newcommand{\nc}{$n\rm_{clusters}$}
\newcommand{\kms}{km s$^{-1}$}
\newcommand*{\myprime}{^{\prime}\mkern-1.2mu}
\newcommand*{\mydprime}{^{\prime\prime}\mkern-1.2mu}

\title[Evaluation of NYMGs]{Evaluation of nearby young moving groups based on unsupervised machine learning}

\author[Lee \& Song]{
Jinhee Lee,$^{1}$\thanks{E-mail: jinhee@uga.edu}
Inseok Song,$^{1}$
\\
$^{1}$Department of Physics and Astronomy, The University of Georgia, Athens, GA 30602\\
}

\date{Accepted XXX. Received YYY; in original form ZZZ}

\pubyear{2019}
\hypersetup{draft}  
\begin{document}
\label{firstpage}
\pagerange{\pageref{firstpage}--\pageref{lastpage}}
\maketitle

\begin{abstract}
Nearby young stellar moving groups have been identified by many research groups with different methods and criteria giving rise to cautions on the reality of some groups.
We aim to utilise moving groups in an unbiased way to create a list of unambiguously recognisable moving groups and their members.
For the analysis, two unsupervised machine learning algorithms (K-means and Agglomerative Clustering) are applied to previously known bona fide members of nine moving groups from our previous study.
As a result of this study, we recovered six previously known groups (AB Doradus, Argus, $\beta$-Pic, Carina, TWA, and Volans-Carina).
Three the other known groups are recognised as well; however, they are combined into two new separate groups (ThOr+Columba and TucHor+Columba).
\end{abstract}




\section{Introduction}

Nearby young stellar moving groups (NYMGs, hereafter) are gravitationally unbound loose stellar associations, and in this paper we focus primarily on NYMGs that are younger than 100 Myr and have mean distances smaller than 100 pc of the Sun.
The first NYMG, the TW Hya association (TWA), was discovered in late 1990 \citep{kas97}.
Then, new NYMGs were identified by discoveries of co-moving groups of stars in a small region of space (within few tens of pc) with signs of similar ages. 
Even though NYMGs have a relatively short history (about 20 years), many NYMGs ($N\sim$10) and their members ($N>$2000) have been discovered (e.g., \citealt{zuc01a, zuc04, tor08, mam07, mal13, ell16, rie17, gag18}). 

Recent studies about NYMGs (e.g., \citealt{fah18, gag18, lee19, zuc19}) recognise nine NYMGs: TW Hya association (TWA), $\beta$-Pic moving group (BPMG), 32 Ori association (ThOr), Tucana-Horologium association (TucHor), Carina association (Carina), Columba association (Columba), Argus association (Argus), AB Doradus moving group (ABDor), and Volans-Carina association (VCA).
Other NYMGs have been claimed (e.g., GAYA 1 \& 2 and AnA by \citealt{tor03}, Carina-Vela by \citealt{mak00}), and they were later redefined or disappeared.
Tucana association \citep{zuc00} and Horologium association \citep{tor00} were merged as a single group by \citet{zuc01b} [Tucana Horologium Association (TucHor)]. 
 \citet{liu15} points out that several moving groups may be merged, rejected, or new groups can be discovered. 
 
In this study, we re-evaluate the well-known nine NYMGs without relying on any previous knowledge.
For this task, the cluster analysis would be an ideal technique.
The cluster analysis is a branch of unsupervised machine learning that finds groups within a dataset without prior reference to the outcome.
For identifying a star belonging to a certain NYMG, the required information is Galactic position ($XYZ$), Galactic velocity ($UVW$), and age. 
Stars occupying a similar space in the age-spatio-kinematic space (i.e., sharing characteristics in the 7D space) can be classified as the same group members.
On the other hand, a star located far from this group of stars in the 7D space can be assessed as non-members.

\section{Method}

\defcitealias{lee19}{Paper I}

\subsection{Data}
As input data, we use bona fide members of nine NYMGs from Lee \& Song (2019; Paper I), which enable us to evaluate NYMGs by comparing clustering results to previously known NYMGs.
Considered nine groups are TWA, BPMG, ThOr, TucHor, Carina, Columba, Argus, ABDor, and VCA.
While the stars' spatio-kinematic information, $XYZ$ and $UVW$, are straightforward to calculate, age is difficult to obtain and generally has a large uncertainty.
In the age range of NYMGs ($\sim$8 to $\sim$100 Myr), age is evaluated using Li equivalent width, position on colour-magnitude diagrams, near-UV excess, or X-ray brightness (e.g., \citealt{zuc04, sod10, rod11, mal13}).
These bona fide members of nine NYMGs are youth-confirmed members.
Instead of using numerical, continuous ages in Myr, we use categorical ages (i.e., age classes) to ensure more efficient groupings in our cluster analysis.
Using common age-dating methods mentioned previously, we age-dated all input stars ($N$=652) into 5 age classes: 1 for 8$-$10 Myr (TWA), 2 for 12$-$20 Myr (BPMG and ThOr), 3 for 30$-$40 Myr (TucHor, Carina, Columba, and Argus), 4 for 90$-$150 Myr (ABDor and VCA), and 5 for older age (field stars).  

We do not consider chemical abundance or metallicity in our analysis because metallicity has a large uncertainty, and it is known that young nearby stars all should have similar abundance \citep{alm09}.

\subsection{Clustering algorithms}

Clustering algorithms find groups in data in a way that members in the same group are more similar to each other than to those in other groups.
There are various clustering algorithms with different strategies and criteria for finding groups.
In this study, we use two algorithms in the {\sc sklearn} package of {\sc python} \citep{ped11}.
Each algorithm has its own specific parameters.

\subsubsection{K-means}

The K-means algorithm finds groups in data by trying to separate samples into a desired number of groups while minimizing within-cluster sum-of-squares.
Within-cluster sum-of-squares, also known as {\it inertia}, is calculated by following equation.
\begin{equation}
\sum_{k=1}^{k}\sum_{x_i\in C_k} (x_i-\mu_k)^2
\end{equation}
Where $C_k$ is the $k$th cluster, and $\mu_k$ is the centroids (the mean) of $C_k$.
K-means requires the number of clusters (\nc) as an input parameter.

\subsubsection{Agglomerative Clustering}

The Agglomerative Clustering algorithm tries to build a hierarchy of clusters starting with clusters consist of an individual sample.
Then, clusters are successively merged together.
To determine which neighbouring clusters will merge, there are several options of linking criteria (e.g., Ward, maximum, average, and single linkage).
We use Ward in this study.
Ward has a merge strategy minimizing the sum of squared differences within all clusters. 
The number of clusters (\nc) is required as an input parameter similar to the K-means method.

\subsection{Procedure}
\subsubsection{Preprocessing of data}

\begin{figure*}
\includegraphics[width=0.9\textwidth]{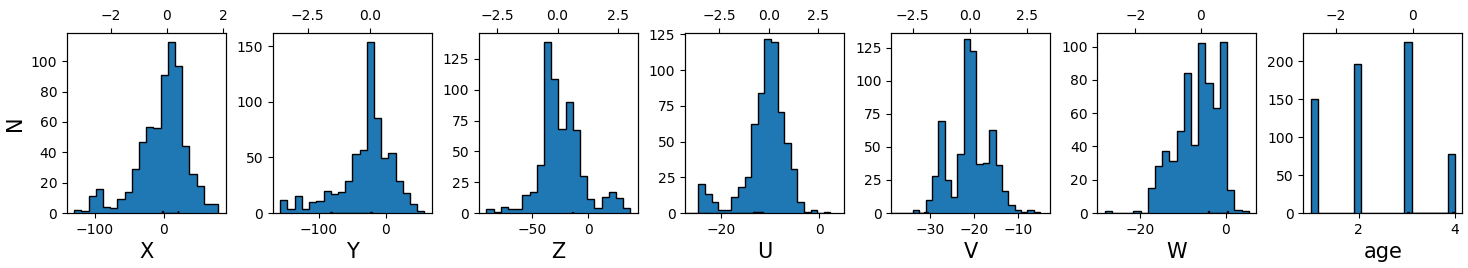}
\caption{Distribution of input data. Raw and transformed scales are displayed along the bottom and the top X-axis, respectively.
Units for raw data are pc ($X$, $Y$, and $Z$) and \kms\ ($U$, $V$, and $W$).
Age (age class) has no unit.}
\label{fig:inputdata}
\end{figure*}

For a reliable clustering, raw data should be transformed to make all variables have a similar range.
Using raw data without a proper transformation can cause biased results.
For example, $XYZ$ values are distributed in a range of 100 (in pc), and $UVW$ values are distributed in a range of 10s (in \kms) (See Fig.~\ref{fig:inputdata}).
Parameters with more extensive ranges can have a more significant effect in clustering. 

Preprocessing in the cluster analysis is a process for homogenising the range of data variables.
There are various transformation functions in the {\sc sklearn} package.
StandardScaler in {\sc sklearn} transforms data to make the distribution has a mean value zero and standard deviation of one, by subtracting the mean and then dividing by the standard deviation of the whole dataset.
Since the outliers have a significant influence on the mean and standard deviation, the StandardScaler would not be a proper choice as a transformation function.
RobustScaler in {\sc sklearn} scales variables using values within percentiles, therefore, is not influenced by a small number of outliers.
In this study, RobustScaler is applied for transforming $XYZUVW$ values, resulting in the transformed variables having a range of approximately -2.5 to +2.5. 
After the transformation, two outliers ($\sim$15$\sigma$ in $U$ and $\sim$7$\sigma$ in $V$) are removed.

The age class has values in the range from +1 to +5.
To make age class spanning the similar range to the transformed $XYZUVW$ variables, we subtract 3.125 to the age class and multiply by 1.25.
This procedure as shown in Fig.~\ref{fig:inputdata} is to ensure that all variables are transformed to match similar scales. \newline

\subsubsection{Strategy for evaluating NYMGs}

When the entire input data are used, both Agglomerative Clustering and K-means algorithms generate nearly the same results with a given \nc.
To find a general trend of groupings from multiple clustering results, for a given algorithm, we performed one thousand trial runs by varying input data slightly for each run.
In each run, 95 per cent of the input data are randomly selected.
Because of the slight difference in the selected data for each run, a different outcome appears each time.
For each combination of the chosen algorithm (K-means and Agglomerative Clustering) and \nc\ (from 3 to 10), we run the calculations one thousand times.
From the clustering results from a total 16,000 runs, we identify groups.
Then, members of these newly identified groups are tracked, and members consistently  assessed into a group ($>$70 per cent of trial runs between the two algorithms) can be recognised as ``bona fide'' members of the group.

\section{Results}

\subsection{Finding a grouping trend}

The most frequent groupings for each of 16 combinations of algorithms and \nc\ values are found and discussed.
To identify the most dominant grouping with a given \nc\ and an algorithm, we need to identify the most frequent pattern of grouping through the one thousand runs.
The tested input parameter, \nc, is in a range of 3 to 10.
Fig.~\ref{fig:nc9} presents results with \nc=9. 
The results with the other \nc\ values are presented in Appendix A.

In these figures, the X-axis and Y-axis represent the indices of runs and stars, respectively.
In each run, each star's resultant clustering results are displayed as a colour-coded point.
For an easier interpretation, the 2D results are sorted along both the X and Y-axes.
The results along the Y-axis (stellar indices) are sorted first.
Stars are sorted by previously known group membership first for comparing the new clustering results to the previously known grouping.
ABDor members are located at the top (index 0 to $\sim$100), and VCA members are located at the bottom.
Then, the results are sorted along the X-axis.
This process sorts the results placing the similar grouping results together, which enable us to easily identify frequent groupings through the entire set of trial runs.
When a star is not included in the 95 per cent sampling, such cases are represented as white dots.

If one extracts a slice along the Y-axis (i.e., takes results from a single run), stars with the same colour indicate that these stars belong to the same group.
The number of colours in one trial run should be the same as \nc.
However, when we look at the results of the entire one thousand runs, more than \nc\ groupings appear generally.
For example, results from K-means with \nc=5 have six colours.
However, if one takes out any single vertical slice, the number of colours is always five for \nc=5.

\begin{figure*}
    \begin{subfigure}{0.49\textwidth}
    \includegraphics[width=\textwidth, angle=0]{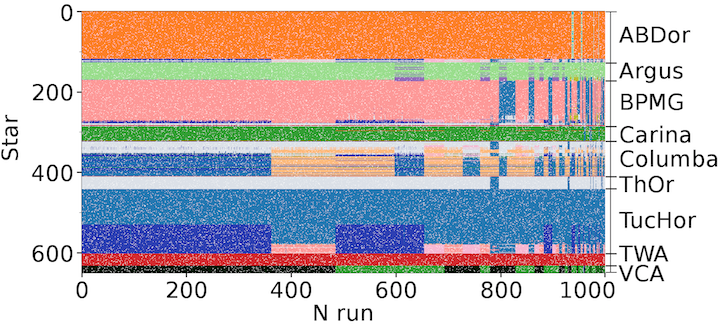}
        \caption{K-means} 
    \end{subfigure}
        \begin{subfigure}{0.49\textwidth}
    \includegraphics[width=\textwidth, angle=0]{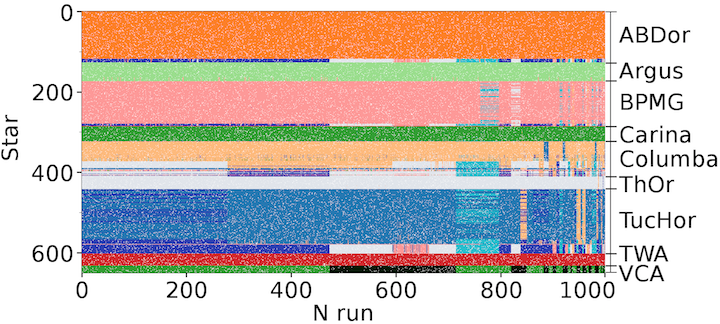}
        \caption{Agglomerative Clustering} 
    \end{subfigure}
\caption{
Results from one thousand runs with \nc=9.  
Left and right panels present the results of K-means and Agglomerative Clustering algorithms, respectively.
The X and Y-axis represent a run number and stellar index, respectively.  
The results are sorted along the X and Y-axis for a better display and interpretation. 
Stars (the Y-axis) are sorted based on previously known grouping.
The ``N run'' (the X-axis) is sorted to locate the most frequent grouping first (at the left side of the result figure).
Therefore, a star placed at 30th row in K-means results might be located at a different row in Agglomerative Clustering results.
A vertical line represents a result from a single run, and stars displayed with the same colour in the line belong to the same group.}    
\label{fig:nc9}
\end{figure*}

\subsection{The best results with \nc=9 and comparison against previously known NYMGs }

\begin{table*}
\begin{threeparttable}
\begin{tabular}{p{0.05\textwidth}p{0.35\textwidth}p{0.05\textwidth}p{0.35\textwidth}p{0.05\textwidth}}
\toprule
\nc\ & \multicolumn{2}{c}{K-means} & \multicolumn{2}{c}{Agglomerative Clustering} \\
\cmidrule(l{2pt}r{2pt}){1-1} \cmidrule(l{2pt}r{2pt}){2-3} \cmidrule(l{2pt}r{2pt}){4-5} 
 & Recognized groups (per cent) & $N\rm_{members}\tnote{a}$   & Recognized groups (per cent) & $N\rm_{members}\tnote{a}$ \\
 \cmidrule(l{2pt}r{2pt}){1-1} \cmidrule(l{2pt}r{2pt}){2-2} \cmidrule(l{2pt}r{2pt}){3-3} \cmidrule(l{2pt}r{2pt}){4-4} \cmidrule(l{2pt}r{2pt}){5-5} 
9 & ABDor (99) & 120 & ABDor (100) & 120 \\
   & Argus (85) & 50 & Argus (100) & 50 \\
   & BPMG (85) & 120 & BPMG (80) & 120 \\
    & TWA (100) & 20 & TWA (100) & 20 \\
   & Carina (70) & 40 & Carina (35) & 40 \\
    & VCA (70) & 20 & VCA (35) & 20 \\
   & ThOr + Columba$\mydprime$ (95) & 50 & ThOr + Columba$\mydprime$ (40) & 50 \\
   & Columba$\myprime$ + TucHor$\myprime$ (60) & 100 & Columba$\myprime$ (80) & 40 \\
   & Columba$\myprime$ + TucHor$\mydprime$ (60) & 70 & TucHor$\myprime$ (50) & 150 \\
\bottomrule
\end{tabular}
\begin{tablenotes}
\item[a] Approximate number of the members. 
\end{tablenotes}
\end{threeparttable}
\caption{The most frequent grouping with \nc=9.
The groupings with other \nc\ values are in Appendix A.
Per cent value indicates that a group is recognised in such per cent of runs.
Groups with a single prime and double prime ($\myprime$ and $\mydprime$) indicate that not the entire but a fraction of group (about 50-80 and 20-50 per cent, respectively) is enclosed to the new group.}
\label{tab:newgrp9}
\end{table*}

The interpretation of clustering results can critically depend on the choice of \nc.
However, choosing a proper \nc\ value is difficult because it requires the acceptance on the number of ``true'' NYMGs.
Nonetheless, because we use nine NYMGs as the input, \nc\ close to 9 would be an appropriate choice to compare our results against the previously known NYMGs.
To evaluate if \nc=``number of input groups'' is a suitable choice, we performed a test with a smaller input data set using only four NYMGs (ABDor, BPMG, TucHor, and TWA).
The two clustering algorithms were utilised in the same way as we used the entire data, but the tested \nc\ range is from 3 to 8.
When \nc=4, both algorithms correctly recovered four NYMGs.
Although this exercise does not guarantee that \nc=9 should be the best choice in considering the entire dataset, the choice of \nc\ may be suitably set to the number of ``expected'' true groupings in the input data.

Table~\ref{tab:newgrp9} and Fig.~\ref{fig:nc9} present the dominant grouping with \nc=9.
Both algorithms find ABDor, Argus, BPMG, and TWA (orange, light green, pink, and red in Fig.~\ref{fig:nc9}, respectively) in more than 80 per cent of cases.
Carina (green) and VCA (dark green) are found with both algorithms when considering a lower percentage (70 per cent with K-means and 35 per cent with Agglomerative Clustering).

Interesting results come from the remaining groups (ThOr, Columba, and TucHor).
Both algorithms find a new group consisting of the entire ThOr and a subset of Columba (95 per cent cases with K-means and 40 per cent cases with Agglomerative Clustering).
With the remaining members, while Agglomerative Clustering splits TucHor (blue) and the subset of Columba (light orange) group, K-means algorithm makes two groups consisting the mixture of TucHor and Columba (blue and dark blue).

Based on the results, with \nc=9, six groups matching to previously known NYMGs are recognised (ABDor, Argus, BPMG, TWA, Carina, and VCA).
A new group containing the entire ThOr and a subset of Columba members is frequently recognised. 
We suggest calling this group ThOr-Col.
The remaining stars are mainly consisting of the entire TucHor and a majority of Columba.
While each K-means and  Agglomerative Clustering algorithm recognises two groups in the remaining members, the two groups from these algorithms do not match well (i.e., the specific content of two groups from both algorithms are different).
Therefore, we define a single group enclosing these two group members.
This group is called THC (TucHor-Col).
Even though groups are taken from results with \nc=9, the number of recognised groups is eight. 

As explained in Section 2.3.2, stars consistently (i.e., $>$70 per cent) assigned to the same group from one thousand runs can be regarded as ``reliable'' members of the group.
From two such lists of members of two algorithms, a set of common members are selected as bona fide members of the newly identified eight NYMGs.
The number of bona fide members listed in Table~\ref{tab:memb} appears to be considerably larger than those in BANYAN series \citep{mal13, gag14, gag18}.
The bona fide members referred in those papers are ``classical'' members to build kinematic models for identifying new members.
Our input data includes not only these classical members but also newly identified  members from the recent studies (See Paper I).

\begin{figure*}
 \includegraphics[width=0.8\textwidth, angle=90]{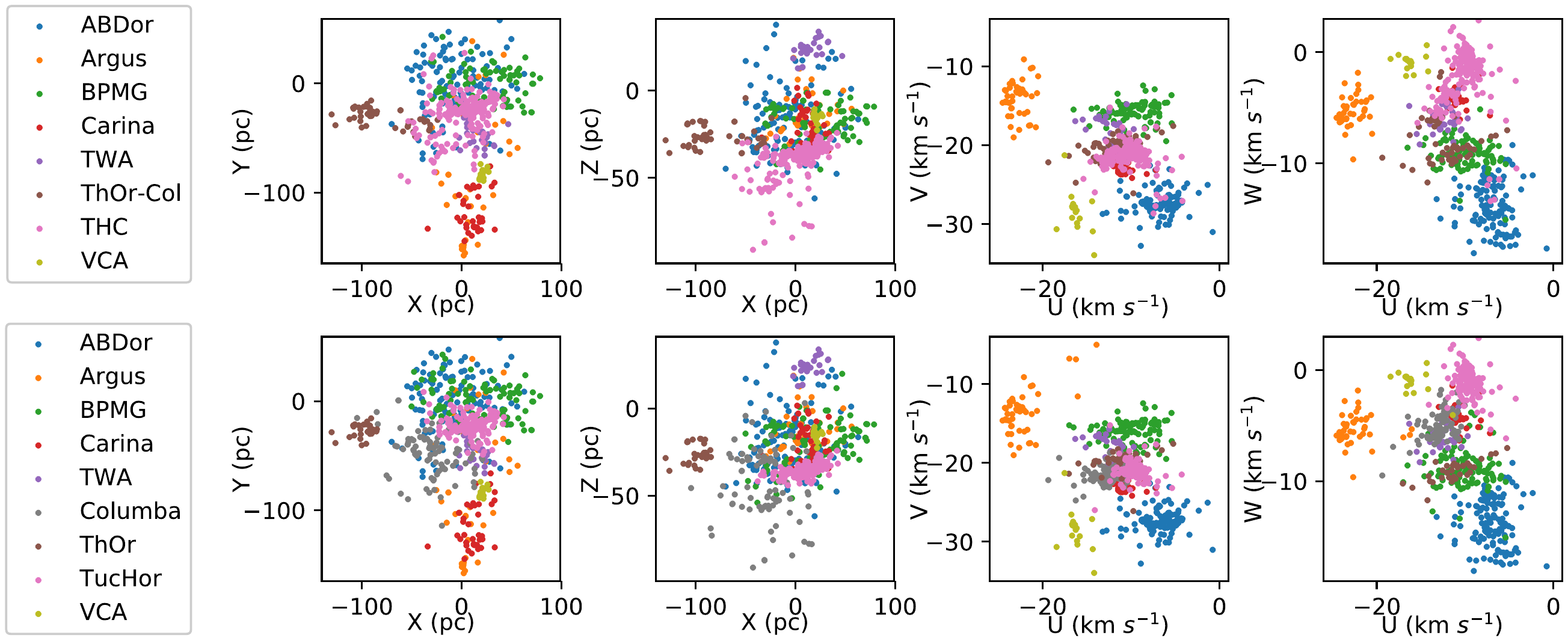}
\caption{Distribution of bona fide members of the new eight group (this study; top) and those of classical groups (input data; bottom).}
\label{fig:grp}
\end{figure*}

Fig.~\ref{fig:grp} compares the distribution of bona fide members of the new eight groups (this study) and those of classical groups (the input data) in the spatio-kinematic space.
As explained earlier, THC and ThOr-Col enclose TucHor, Columba, and ThOr, while the other six groups are recovered almost the same to the previous NYMGs.

\begin{table}
\footnotesize
\begin{tabular}{p{0.035\textwidth}p{0.15\textwidth}p{0.025\textwidth}p{0.08\textwidth}p{0.08\textwidth}}
\toprule
Group & Name & SpT & R.A. (hh:mm:ss) & Dec. (dd:mm:ss) \\ \hline
     ABDor &                    PW And &       K0Ve &     00:18:20.89 &     +30:57:22.1 \\ 
     ABDor &   2MASS J00192626+4614078 &         M8 &     00:19:26.27 &     +46:14:07.8 \\ 
     ABDor &       2MASS J00489+4435AB &         M4 &     00:48:58.18 &     +44:35:08.8 \\ 
\multicolumn{5}{c}{\dots} \\ 
\multicolumn{5}{c}{\dots} \\ 
\multicolumn{5}{c}{\dots} \\      
     Argus &                CD-29 2360 &       K3Ve &     05:34:59.23 &     -29:54:04.0 \\ 
     Argus &                CD-42 2906 &        K1V &     07:01:53.41 &     -42:27:56.2 \\ 
     Argus &                CD-48 2972 &        G8V &     07:28:22.03 &     -49:08:37.7 \\ 
\multicolumn{5}{c}{\dots} \\ 
\multicolumn{5}{c}{\dots} \\ 
\multicolumn{5}{c}{\dots} \\ 
\bottomrule
\end{tabular}
\caption{Sample bona fide members of the newly defined groups.
The full data are available online.}
\label{tab:memb}
\end{table}

\section{Discussion and Conclusion}

In this study, we carried out a pattern recognition in the age-spatio-kinematic space for an unbiased evaluation of NYMGs using an unsupervised machine learning method.
The grouping was evaluated by numerous trials of the clustering algorithms with 95 per cent random sampling of the previously known NYMG members.
The most frequent grouping pattern was found with a given \nc.
While there are some differences between results with K-means and Agglomerative Clustering algorithms, there is a significant similarity between these results.

ABDor is the first recognised group using both algorithms at \nc=3, and Argus is the second recognised group (\nc=4).
Carina and VCA tend to form a single group at a small \nc, which indicates that these two groups share similar characteristics ($XYZUVW$ and age).  
However, with a large \nc\ value (\nc=9$-$10), these two groups are clearly separated as previously known groupings (i.e., recognised as Carina and VCA).
This implies that these two groups share similar characteristics but probably not a single entity.
Similar to the case of Carina and VCA, BPMG and TWA have similar age-spatio-kinematic characteristics, making them into a single entity with a smaller \nc\ value ($<$7). 
However, these two groups are split clearly with larger \nc\ values.

One interesting result comes from Columba and ThOr.
Both algorithms create a merged group, ThOr-Col.
The entire ThOr and majority of Columba members form this new group.
THC group consists of almost the entire TucHor and a half of Columba members.
\citet{zuc11} proposed to combine members of TucHor and Columba, because there is a difficulty assigning some of them into either TucHor of Columba.
Our result supports the claim of \citet{zuc11}.

\begin{table}
\small
\begin{threeparttable}
\begin{tabular}{lll}
\toprule
Name & New group\tnote{a} & Old group (prob)\tnote{b} \\ \hline
1RXS J235133.3+312720 (AB) & ABDor & ABDor (100) \\
 2MASS J02192210-3925225B & THC & TucHor (100) \\
 TWA 27 & TWA & TWA (100) \\
2MASS J00413538-5621127 & THC & TucHor (100) \\
   2MASS J04373613-0228248 & BPMG & BPMG (100) \\
   CD-35 2722 & ABDor & ABDor (100) \\
   CD-52 381 & THC & Columba (99) \\
   2MASS J01123504+1703557 & ABDor & ABDor (100) \\
  TWA 5A & TWA & TWA (100) \\
  $\beta$-Pic & BPMG & BPMG (100) \\
   $\eta$ Tel & BPMG & BPMG (97) \\
           \hline 
2MASS J01033563-5515561 & $-$\tnote{c} & TucHor (100) \\
TYC 9486-927-1 & $-$\tnote{c} & BPMG (100) \\
 2MASS J22501512+2325342A  & $-$\tnote{c} & ABDor (100) \\
 AB Pic & $-$\tnote{c} & Columba (100) \\
  2MASS J21140802-2251358 & $-$\tnote{c} & BPMG (100) \\
  PZ Tel & $-$\tnote{c} & BPMG (100) \\
 SDSS J111010.01+011613.1  & $-$\tnote{c} & ABDor (100) \\
\bottomrule
\end{tabular}
\begin{tablenotes}
\item[a] Assigned group in this study.
\item[b] Previously known groups in Paper I with a  Bayesian membership probability (per cent).
\item[c] These stars are in the input data, but not retained as bona fide members in Section 3.2.
\end{tablenotes}
\end{threeparttable}
\caption{Membership of planet-host stars.
While bona fide members based on Bayesian membership probability in Paper I are used as input data, there are members not retained as bona fide members in cluster analysis.}
\label{tab:planet}
\end{table}

The input data include eighteen planet-host stars.
Among these 18 stars, 11 stars are retained as bona fide members via cluster analysis in this study, while the other seven stars are not consistently recognised as members of the newly defined groups.
Table~\ref{tab:planet} shows membership of these stars.
Membership from this study and Bayesian membership probability from Paper I are compared.
In the case of eleven retained bona fide members, their memberships are consistent with those in Paper I. 

In this study, previously known NYMGs are evaluated using previously known bona fide members of the groups.
If old field stars are included in the analysis, the results will be different.
For example, ABDor and Argus are easily separated from other groups, because of their unique $XYZ$ and $UVW$.
However, because the two groups are relatively old themselves, any inclusion of field stars in the analysis as well will serve to undermine the appearance of separation.
The number density of NYMGs is lower than that of field stars, and there are many field stars sharing similar $XYZ$ and $UVW$ with ABDor and Argus members.

Expanding this study, young nearby stars and field stars should be analysed without knowledge about known NYMGs for the most unbiased and objective analysis of nearby stars.
The new configuration of groups will provide more reliable and objective investigations of young nearby stars and the environment of the solar neighbourhood.

\section*{Acknowledgements}
We acknowledge the important and critical review by the referee that improved the manuscript significantly.
We also thank Tara Cotten for valuable comments and suggestions.







\section{SUPPORTING INFORMATION}
Additional supporting information may be found in the online version of this article. \newline
Appendix A: results of K-means and Agglomerative Clustering with several \nc\ values.



\null\newpage

\appendix

\onecolumn

\section{Results of K-means and Agglomerative Clustering with several \nc\ values}

\begin{figure*}
    \begin{subfigure}{0.49\textwidth}
    \includegraphics[width=\textwidth, angle=0]{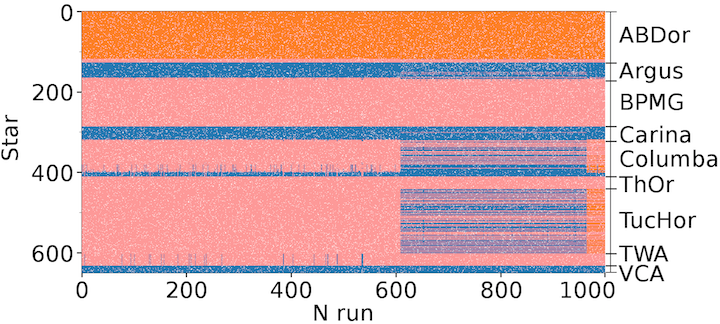}
        \caption{K-means} 
    \end{subfigure}
        \begin{subfigure}{0.49\textwidth}
    \includegraphics[width=\textwidth, angle=0]{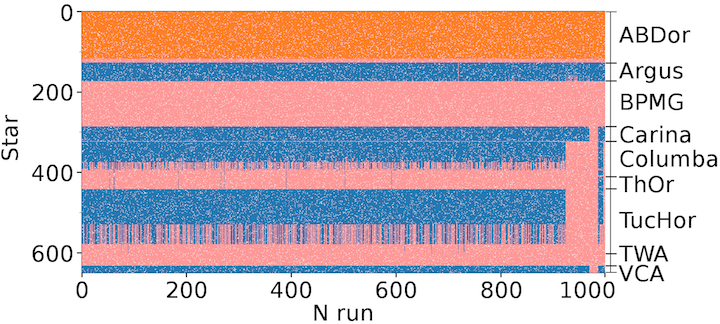}
        \caption{Agglomerative Clustering} 
    \end{subfigure}
\caption{Results from one thousand runs with \nc=3.  
Left and right panels present the results of K-means and Agglomerative Clustering algorithms, respectively.
The X and Y-axis represent a run number and stellar index, respectively.  
The results are sorted along the X and Y-axis for a better display and interpretation. 
Stars (the Y-axis) are sorted based on previously known grouping.
The ``N run'' (the X-axis) is sorted to locate the most frequent grouping first (at the left side of the result figure).
Therefore, a star placed at 30th row in K-means results might be located at a different row in Agglomerative Clustering results.
A vertical line represents a result from a single run, and stars displayed with the same colour in the line belong to the same group.
This figure is available online.}    
\label{fig:nc3}
\end{figure*}

\begin{figure*}
    \begin{subfigure}{0.49\textwidth}
    \includegraphics[width=\textwidth, angle=0]{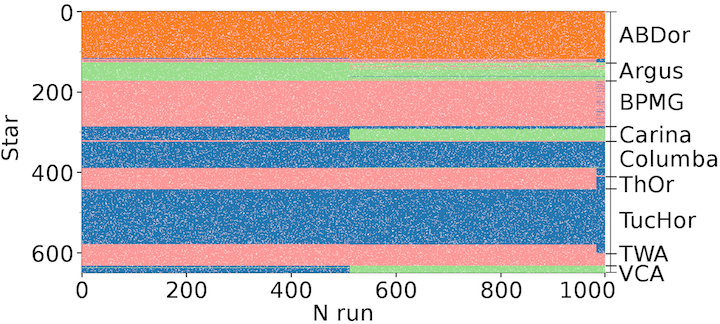}
        \caption{K-means} 
    \end{subfigure}
        \begin{subfigure}{0.49\textwidth}
    \includegraphics[width=\textwidth, angle=0]{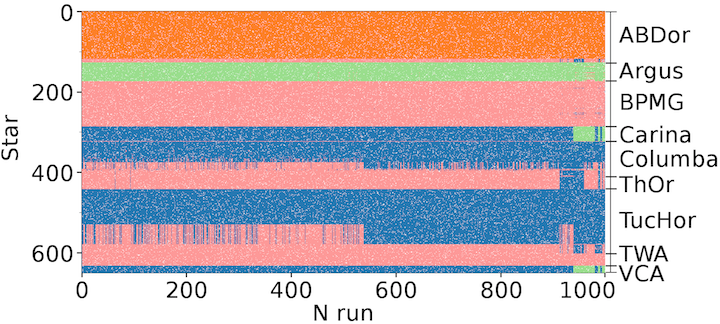}
        \caption{Agglomerative Clustering} 
    \end{subfigure}
\caption{Same to Fig~\ref{fig:nc3} but for \nc=4.}    
\label{fig:nc4}
\end{figure*}

\begin{figure*}
    \begin{subfigure}{0.49\textwidth}
    \includegraphics[width=\textwidth, angle=0]{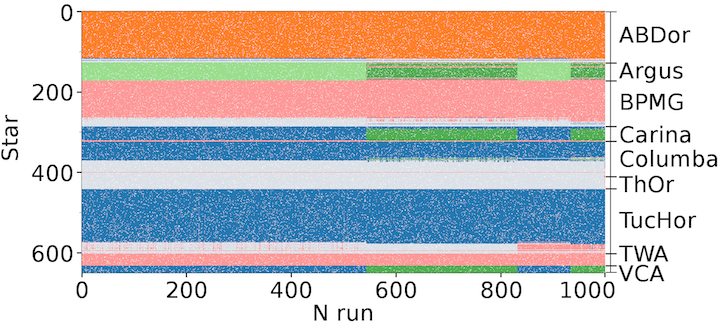}
        \caption{K-means} 
    \end{subfigure}
        \begin{subfigure}{0.49\textwidth}
    \includegraphics[width=\textwidth, angle=0]{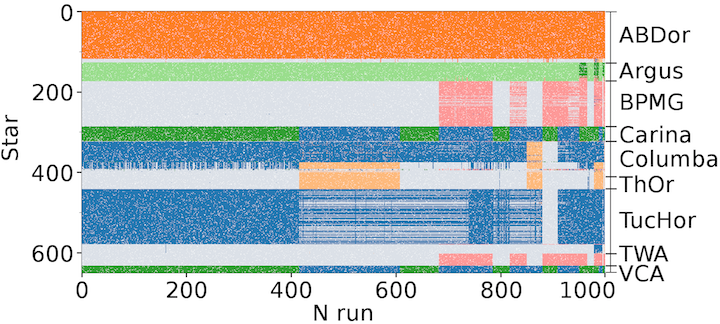}
        \caption{Agglomerative Clustering} 
    \end{subfigure}
\caption{Same to Fig.~\ref{fig:nc3} but for \nc=5. }    
\label{fig:nc5}
\end{figure*}

\begin{figure*}
    \begin{subfigure}{0.49\textwidth}
    \includegraphics[width=\textwidth, angle=0]{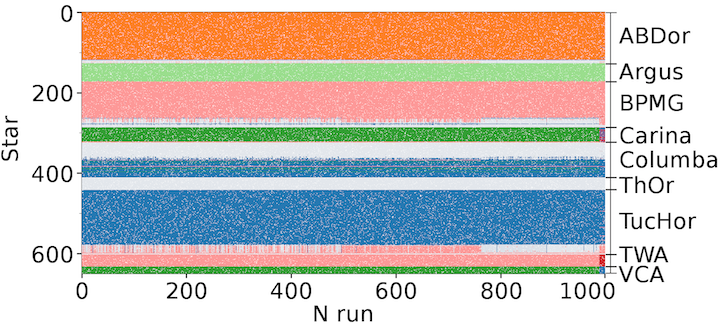}
        \caption{K-means} 
    \end{subfigure}
        \begin{subfigure}{0.49\textwidth}
    \includegraphics[width=\textwidth, angle=0]{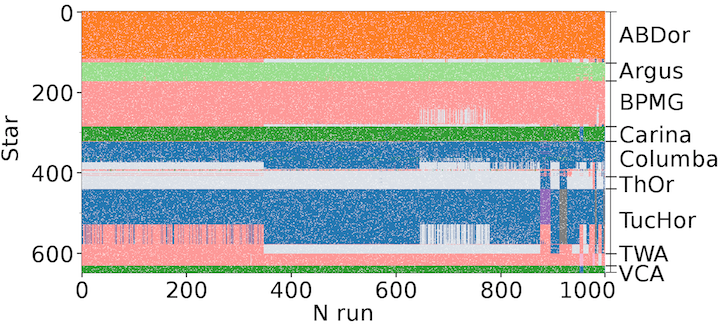}
        \caption{Agglomerative Clustering} 
    \end{subfigure}
\caption{Same to Fig.~\ref{fig:nc3} but for \nc=6.}    
\label{fig:nc6}
\end{figure*}

\begin{figure*}
    \begin{subfigure}{0.49\textwidth}
    \includegraphics[width=\textwidth, angle=0]{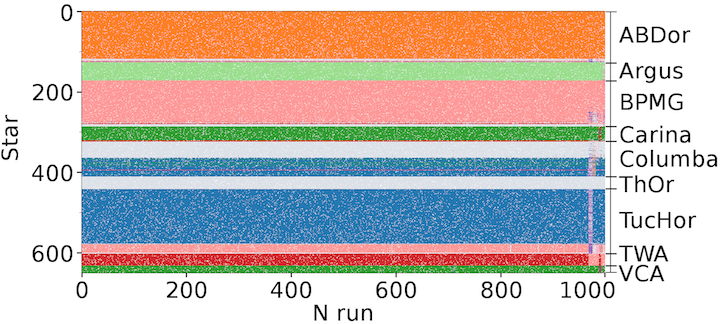}
        \caption{K-means} 
    \end{subfigure}
        \begin{subfigure}{0.49\textwidth}
    \includegraphics[width=\textwidth, angle=0]{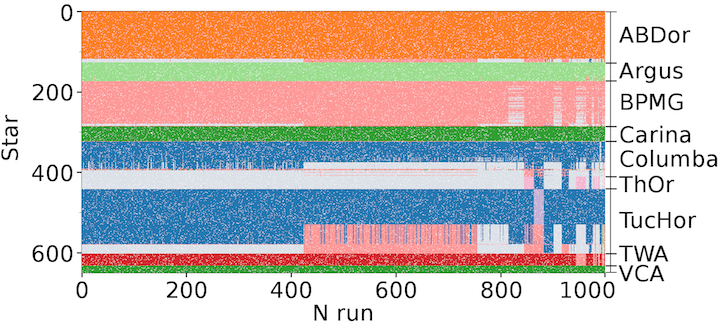}
        \caption{Agglomerative Clustering} 
    \end{subfigure}
\caption{Same to Fig.~\ref{fig:nc3} but for \nc=7.}    
\label{fig:nc7}
\end{figure*}

\begin{figure*}
    \begin{subfigure}{0.49\textwidth}
    \includegraphics[width=\textwidth, angle=0]{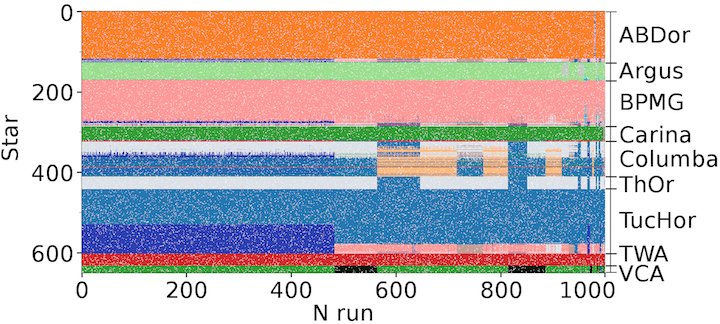}
        \caption{K-means} 
    \end{subfigure}
        \begin{subfigure}{0.49\textwidth}
    \includegraphics[width=\textwidth, angle=0]{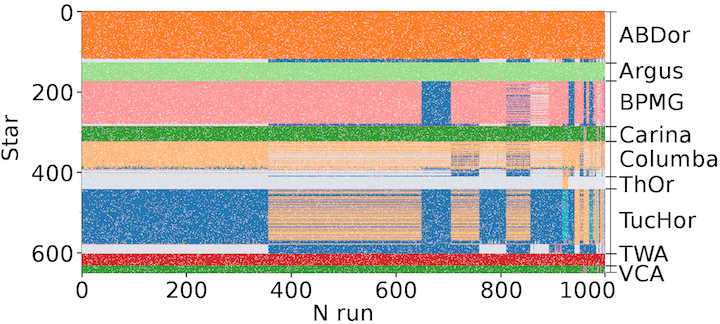}
        \caption{Agglomerative Clustering} 
    \end{subfigure}
\caption{Same to Fig.~\ref{fig:nc3} but for \nc=8.}    
\label{fig:nc8}
\end{figure*}

\begin{figure*}
    \begin{subfigure}{0.49\textwidth}
    \includegraphics[width=\textwidth, angle=0]{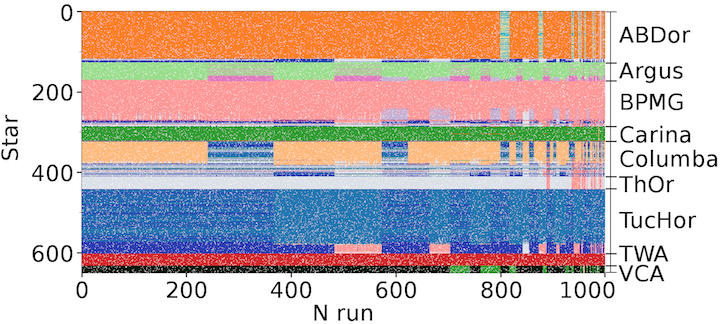}
        \caption{K-means} 
    \end{subfigure}
        \begin{subfigure}{0.49\textwidth}
    \includegraphics[width=\textwidth, angle=0]{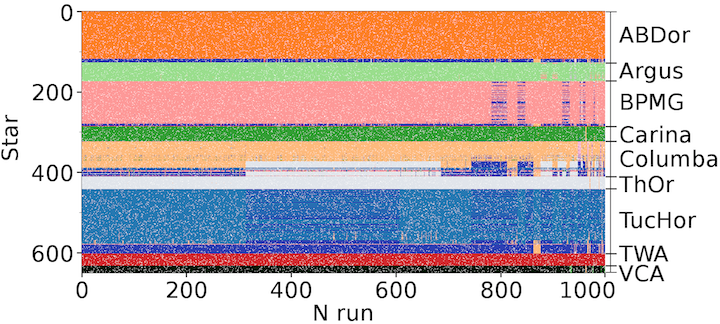}
        \caption{Agglomerative Clustering} 
    \end{subfigure}
\caption{Same to Fig.~\ref{fig:nc3} but for \nc=10.}    
\label{fig:nc10}
\end{figure*}

The grouping trend with a given \nc\ is presented in Figs.~\ref{fig:nc3}-~\ref{fig:nc10} and Tabel~\ref{tab:newgrp}, and discussed here.

\begin{table*}
\begin{threeparttable}
\begin{tabular}{p{0.05\textwidth}p{0.35\textwidth}p{0.05\textwidth}p{0.35\textwidth}p{0.05\textwidth}}
\toprule
\nc\ & \multicolumn{2}{c}{K-means} & \multicolumn{2}{c}{Agglomerative Clustering} \\
\cmidrule(l{2pt}r{2pt}){1-1} \cmidrule(l{2pt}r{2pt}){2-3} \cmidrule(l{2pt}r{2pt}){4-5} 
 & Recognized groups (per cent) & $N\rm_{members}\tnote{a}$   & Recognized groups (per cent) & $N\rm_{members}\tnote{a}$ \\
 \cmidrule(l{2pt}r{2pt}){1-1} \cmidrule(l{2pt}r{2pt}){2-2} \cmidrule(l{2pt}r{2pt}){3-3} \cmidrule(l{2pt}r{2pt}){4-4} \cmidrule(l{2pt}r{2pt}){5-5} 
3  & ABDor (95)          & 120 & ABDor (100) & 120 \\
    & BPMG + TWA + ThOr + TucHor+ Columba$\myprime$  (60) &430 & BPMG + TWA + ThOr + TucHor$\mydprime$ + Columba$\mydprime$ (93) & 200 \\  
   & Argus + Carina+ VCA  (60) & 100 & Argus + Carina + VCA + TucHor$\myprime$ + Columba$\myprime$ (93) & 330 \\
\hline
4 & ABDor  (100) & 120 & ABDor (100) & 120 \\
   & Argus (50) & 50 & Argus (95) & 50 \\
   & BPMG + ThOr + TWA + Columba$\mydprime$ + TucHor$\mydprime$ (98) & 200 & BPMG + ThOr + TWA + Columba$\mydprime$ + TucHor$\myprime$ (55) & 280 \\
   & Carina + VCA + Columba$\myprime$ + TucHor$\myprime$ (50) & 280 &  Carina + VCA + Columba$\myprime$ + TucHor$\myprime$ (55) & 200 \\
\hline   
5 & ABDor (100) & 120 & ABDor (100) &120 \\
   & Argus (63) & 50 & Argus (98) & 50 \\
   & TWA + BPMG$\myprime$ (85) & 130 & BPMG + ThOr + TWA + Columba$\mydprime$ + TucHor$\mydprime$ (50) & 200 \\
   & ThOr + Columba$\myprime$ + BPMG$\mydprime$ + TucHor$\mydprime$ (85) & 100 & Carina + VCA (60) & 50 \\
   & Carina + VCA + TucHor$\myprime$ + Columba$\myprime$ (63) & 250 & TucHor + Columba$\myprime$ (45) & 200 \\
\hline
6 & ABDor (100) & 120 & ABDor (100) & 120 \\
   & Argus (100) & 50 & Argus (97) & 50 \\
   & Carina + VCA (98) & 50 & Carina + VCA (99) & 50 \\
   & BPMG$\myprime$ + TWA (80) & 130 & BPMG + TWA + TucHor$\myprime$ (40) & 200 \\
   & ThOr + Columba$\myprime$ + BPMG$\mydprime$ (100) & 100 & ThOr +  Columba$\mydprime$ + TucHor$\mydprime$ (55) & 80  \\
   & TucHor$\myprime$ + Columba$\myprime$ (100) & 200 &   TucHor$\myprime$ + Columba$\myprime$ (90) & 150 \\
\hline
7 & ABDor (100) & 120 & ABDor (100) & 120 \\
   & Argus (99) & 50 & Argus (100) & 50 \\
   & Carina + VCA (99) & 50 & Carina + VCA (100) & 50 \\
   & Columba$\myprime$ + TucHor$\myprime$ (99) & 200 & Columba$\myprime$ + TucHor$\myprime$ (95) & 200  \\   
    & ThOr + Columba$\myprime$  (99) & 80 & ThOr + Columba$\mydprime$ + TucHor$\mydprime$ (60) & 70 \\
    & TWA (95) & 30  & TWA (95) & 30 \\
   & BPMG + TucHor$\mydprime$ (95) & 120 & BPMG (45) & 120 \\
\hline   
8 & ABDor (99) & 120 & ABDor (100) & 120 \\
   & Argus (95) & 50 & Argus (100) & 50 \\
   & BPMG (50) & 120 & BPMG (85) & 120 \\
    & TWA (100) & 30 & TWA (100) & 30 \\ 
   & Carina + VCA (85) & 50 & Carina + VCA (98) & 50 \\
   & ThOr + Columba$\mydprime$ (80) & 100 & ThOr + Columba$\mydprime$ + TucHor$\mydprime$ (45) & 70 \\
   & TucHor$\myprime$ (50) & 70 & Columba$\myprime$ (35) & 70 \\
   & TucHor$\myprime$ + Columba$\myprime$ (50) & 100 & TucHor$\myprime$  (45) & 140 \\
\hline   
10 & ABDor (95) & 120 & ABDor (100) & 120 \\
     & Argus (70) & 50 & Argus (100) & 50 \\
     & BPMG (80) & 120 &  BPMG (90) & 120 \\
     & Carina (80) & 40 & Carina (100) & 40 \\
     & TWA (100) & 20 & TWA (100) & 20 \\
     & VCA (80) & 20 & VCA (100) & 20 \\
     & ThOr + Columba$\mydprime$ (90) & 50    & ThOr + Columba$\myprime$ (45) &70 \\
     & Columba$\myprime$ (80) & 50     & Columba$\myprime$ (100) & 50 \\
     & TucHor$\myprime$ (65) & 80 & TucHor$\myprime$ (50) & 80 \\
     & TucHor$\myprime$ (66) & 80 & TucHor$\myprime$ (50) & 80 \\
\bottomrule
\end{tabular}
\begin{tablenotes}
\item[a] Approximate number of the members.
\end{tablenotes}
\end{threeparttable}
\caption{The most frequent grouping with a given \nc.
Per cent value indicates that a group is recognised in such per cent of runs.
Groups with a single prime and double prime ($\myprime$ and $\mydprime$) indicate that not the entire but a fraction of group (about 50-80 and 20-50 per cent, respectively) is enclosed to the new group.}
\label{tab:newgrp}
\end{table*}

\subsection{\nc=3}

With \nc\ of 3, ABDor is solely identified with both algorithms in more than 95 per cent cases (orange in the figure).
A large group (pink colour with both algorithms) consisting of BPMG, TWA, ThOr, TucHor, a subset of Columba, few ABDor members is recognised with both algorithms with a different number of members. 
The other group (blue) consists of Argus, Carina, VCA, and a subset of Columba.
In the case of Agglomerative Clustering, this group additionally includes about half of TucHor members (i.e., more inclusive than the group from K-means).

\subsection{\nc=4}

Similar to the results with \nc=3, ABDor is solely identified with both algorithms in 100 per cent cases (orange).
K-means algorithm identifies a group presented with pink colour consisting of BPMG, ThOr, TWA, a subset of Columba, a subset of TucHor, and a few ABDor members in $\sim$98 per cent cases.
Argus (light green) is solely identified in 50 per cent cases, and a group presented with blue colour consists of Carina, VCA, and a majority of Columba and TucHor members. 
Agglomerative Clustering makes similar results with those from the K-means algorithm while the number of members is different.

\subsection{\nc=5}

K-means creates results similar to the ones with \nc=4.
The difference is that a group illustrated with pink with \nc=4 is split into two groups; a group displayed with light blue consists of ThOr, a half of Columba, and a subset of BPMG and TucHor.
The other group displayed with pink consists of BPMG, and TWA members. \newline
In case of Agglomerative Clustering, compared to \nc=4, a group displayed with blue split into two groups; a group consisting of Carina and VCA (green) and the other group enclosing a subset of TucHor and Columba (blue).

\subsection{\nc=6}

Both algorithms identify similar grouping.
ABDor and Argus (orange and light green, respectively), a group enclosing Carina and VCA (green), another group including BPMG and TWA (pink), and the other group consisting of ThOr and a subset of Columba (light blue) are recognised with both algorithms.
The remaining members mainly consist of TucHor and Columba members and construct a group (blue).

\subsection{\nc=7}

Compared to \nc=6, both algorithms split TWA (red) from the combined group illustrated with pink in results with \nc=6.
The two algorithms make a slightly different grouping for a group displayed with light blue.
A group illustrated with pink encloses a subset of TucHor with K-means in more than 95 cases, while Agglomerative Clustering does not enclose a subset of TucHor to the group in majority cases ($\sim$60 per cent).

\subsection{\nc=8}

Compared to \nc=7, K-means splits a group illustrated with blue into two groups (blue and dark blue).
Each group consists of a mixture of Columba and TucHor.
In the case of Agglomerative Clustering, a group presented with blue with \nc=7 is split into two groups consisting of Columba (light orange) and TucHor (blue).

\subsection{\nc$\geq$10}

In addition to recognising all previously known moving groups, both K-means and Agglomerative tends to split TucHor into two groups (blue and dark blue).
When \nc\ is eleven, BPMG or ABDor begin to be split.
Both BPMG and ABDor are well-known single groups, and we postulate that these results cannot be physically true.
The larger \nc\ values divide more groups as expected and get more complicated to interpret. 
Therefore, we do not elaborate on the analysis of results with \nc\ larger than 10. 

\bsp	
\label{lastpage}
\end{document}